%% file: DESY-03-218.tex
\documentclass[zpreprint,zbstdefault]{./LaTeX/zeus/zeus_paper}
\usepackage[english]{babel}
\input ./LaTeX/zeus/zeus_def.tex
\begin{document}
%
%
\include{DESY-03-218-tit}

\include{DESY-03-218-aut}
%
\include{DESY-03-218-txt}
\include{DESY-03-218-ref}
\include{DESY-03-218-tab}

\include{DESY-03-218-fig}

%
%
\end{document}

%% file: LaTeX/zeus/zeus_def.tex
\newcommand{\DO}{{D{\O}}\xspace}

\chardef\usc=95
\chardef\til=126
\catcode`\@=11 
\DeclareRobustCommand\xdotspace{\futurelet\@let@token\@xdotspace}
\def\@xdotspace{%
  \ifx\@let@token.\else
  \ifx\@let@token\bgroup.\else
  \ifx\@let@token\egroup.\else
  \ifx\@let@token\/.\else
  \ifx\@let@token\ .\else
  \ifx\@let@token~.\else
  \ifx\@let@token!.\else
  \ifx\@let@token,.\else
  \ifx\@let@token:.\else
  \ifx\@let@token;.\else
  \ifx\@let@token?.\else
  \ifx\@let@token/.\else
  \ifx\@let@token'.\else
  \ifx\@let@token).\else
  \ifx\@let@token-.\else
  \ifx\@let@token\@xobeysp.\else
  \ifx\@let@token\space.\else
  \ifx\@let@token\@sptoken.\else
   .\space
   \fi\fi\fi\fi\fi\fi\fi\fi\fi\fi\fi\fi\fi\fi\fi\fi\fi\fi}
\catcode`\@=12 
\newcommand{\CL}[1]{$#1\%$~C.L\xdotspace}
\newcommand{\stru}[2]{%
   \relax\ifmmode\hbox{\vrule height#1 depth#2 width0pt}%
   \else\vrule height#1 depth#2 width0pt\fi}

\newcommand{\Ronum}[1]{\uppercase\expandafter{\romannumeral#1}}
\newcommand{\ronum}[1]{\expandafter{\romannumeral#1}}
\DeclareRobustCommand{\LaTeXZ}{%
  \LaTeX\kern-.05em4\kern-.1em
  {\raisebox{-0.2ex}{$\scriptstyle\text{ZEUS}$}}\xspace}

\newcommand{\eq}[1]{(\ref{eq-#1})}

\newcommand{\fig}[1]{Fig.~\ref{fig-#1}}

\newcommand{\tab}[1]{Table~\ref{tab-#1}}

\DeclareMathAlphabet{\mathbf}{OT1}{cmr}{bx}{sl}
\newcommand{\eVdist}{\kern-0.06667em}

\newcommand{\gev}{{\,\text{Ge}\eVdist\text{V\/}}}
\newcommand{\tev}{{\,\text{Te}\eVdist\text{V\/}}}

\newcommand{\pbi}{\,\text{pb}^{-1}}

\newcommand{\mm}{\,\text{mm}}
\newcommand{\cm}{\,\text{cm}}

\newcommand{\slashfrac}[2]{%
  \raisebox{0.5ex}{\ensuremath #1}\kern-0.12em/\kern-0.08em
  \raisebox{-.8ex}{\ensuremath #2}}

\newcommand{\sqr}[3]{%
    {\vcenter{\hrule height.#3ex\hbox{\vrule width.#2ex height#1ex
     \kern#1ex\vrule width.#3ex}\hrule height.#2ex}}}

\newcommand{\widebar}[1]{%
   \mkern1.5mu\overline{\mkern-1.5mu#1\mkern-1.mu}\mkern1.mu}
\catcode`\@=11 
\newcommand{\parenbar}{\mathpalette\p@renb@r}
\def\p@renb@r#1#2{\vbox{%
  \ifx#1\scriptscriptstyle \dimen@.7em\dimen@ii.2em\else
  \ifx#1\scriptstyle \dimen@.8em\dimen@ii.25em\else
  \dimen@1em\dimen@ii.4em\fi\fi \offinterlineskip
  \ialign{\hfill##\hfill\cr
    \vbox{\hrule width\dimen@ii}\cr
    \noalign{\vskip-.3ex}%
    \hbox to\dimen@{$\mathchar300\hfil\mathchar301$}\cr
    \noalign{\vskip-.3ex}%
    $#1#2$\cr}}}
\catcode`\@=12 

\newcommand{\qbar}{\widebar{q}}

\newcommand{\sthr}{{\scriptscriptstyle 3}}

\newcommand{\ctws}{\cos^2\theta_W}

\newcommand{\stws}{\sin^2\theta_W}

\newcommand{\tw}{\theta_W}

\newcommand{\IP}{{\rm I$\kern-0.01667em$P}\xspace}

\newcommand{\NC}{{\rm NC}}

\mathchardef\qsm=63
\mathchardef\pls=43
\mathchardef\mns=512
\mathchardef\plm=518
\mathchardef\eql=61
\mathchardef\smallleft=300
\mathchardef\smallright=301
\mathchardef\les=316
\mathchardef\gre=318
\mathchardef\leq=532
\mathchardef\grq=533

\catcode`\@=11 
\newcounter{pict@width}
\newcounter{pict@height}
\newlength{\pict@scale}
\newcommand{\psfigadd}[4]{%
\setcounter{pict@width}{1*\ratio{#2+\pict@scale/2}{\pict@scale}}
\setcounter{pict@height}{1*\ratio{#3+\pict@scale/2}{\pict@scale}}
\setlength{\unitlength}{\pict@scale}
\hbox to #2{\hspace{-\fill}\begin{picture}(\thepict@width,\thepict@height)
\put(0,0){\psfig{figure=#1,width=#2,height=#3,clip=}}
\SetScale{0.283466457}
\SetWidth{1.763889}
{#4}
\end{picture}}
}
\newcounter{pict@widthfst}
\newcounter{pict@widthscd}
\newcounter{pict@widthtot}
\newcommand{\psfigaddtwo}[7]{%
\setcounter{pict@widthfst}{1*\ratio{#2+\pict@scale/2}{\pict@scale}}
\setcounter{pict@widthscd}{1*\ratio{#2+#4+\pict@scale/2}{\pict@scale}}
\setcounter{pict@widthtot}{1*\ratio{#2+#4+#6+\pict@scale/2}{\pict@scale}}
\setcounter{pict@height}{1*\ratio{#3+\pict@scale/2}{\pict@scale}}
\setlength{\unitlength}{\pict@scale}
\hbox{\hspace{-\fill}\begin{picture}(\thepict@widthtot,\thepict@height)
\put(0,0){\psfig{figure=#1,width=#2,height=#3,clip=}}
\put(\thepict@widthscd,0){\psfig{figure=#5,width=#6,height=#3,clip=}}
\SetScale{0.283466457}
\SetWidth{1.763889}
{#7}
\end{picture}}
}
\newcommand{\psfigror}[4]{%
\setcounter{pict@width}{1*\ratio{#2+\pict@scale/2}{\pict@scale}}
\setcounter{pict@height}{1*\ratio{#3+\pict@scale/2}{\pict@scale}}
\setlength{\unitlength}{\pict@scale}
\hbox{\begin{picture}(\thepict@width,\thepict@height)
\put(0,\thepict@height){\psfig{figure=#1,width=#3,height=#2,clip=,angle=270}}
\SetScale{0.283466457}
\SetWidth{1.763889}
{#4}
\end{picture}}
}
\newcommand{\psfigrol}[4]{%
\setcounter{pict@width}{1*\ratio{#2+\pict@scale/2}{\pict@scale}}
\setcounter{pict@height}{1*\ratio{#3+\pict@scale/2}{\pict@scale}}
\setlength{\unitlength}{\pict@scale}
\hbox{\begin{picture}(\thepict@width,\thepict@height)
\put(0,0){\psfig{figure=#1,width=#3,height=#2,clip=,angle=90}}
\SetScale{0.283466457}
\SetWidth{1.763889}
{#4}
\end{picture}}
}
\catcode`\@=12 
\newlength\listtextwidth

\catcode`\@=11 
\newlength{\@tabfninsert}
\newlength{\@tabfnwidth}
\newcommand{\tabfootnote}[2]{%
  \setlength{\@tabfninsert}{0.8em}
  \setlength{\@tabfnwidth}{\textwidth}
  \addtolength{\@tabfnwidth}{-\@tabfninsert}
  \addtolength{\@tabfnwidth}{-0.4em}
  \noindent\makebox[\@tabfninsert][r]{\footnotesize$^{#1}$\hfil}\hfill%
  \parbox[t]{\@tabfnwidth}{\footnotesize #2\hfill}}
\catcode`\@=12 

%% file: DESY-03-218-tit.tex
\prepnum{{DESY--03--218}}
\date{December 2003}

\title{
      Search for contact interactions, large extra dimensions and 
      finite quark radius in $ep$ collisions at HERA
      }                                                       
                    
\author{ZEUS Collaboration}

\abstract{
    A search for physics beyond the Standard Model
    has been performed with high-$Q^2$ neutral current deep inelastic 
    scattering events recorded with the ZEUS detector at HERA.
    Two data sets, 
    $e^{+} p \rightarrow e^{+} X$ and $e^{-}p  \rightarrow e^{-} X$, 
    with respective integrated luminosities of $112\pbi$ and $16\pbi$,
    were analyzed.  
    The data reach $Q^2$ values as high as  $40\:000\gev^2$.
    No significant deviations 
    from Standard Model predictions were observed.  Limits 
    were derived on the effective mass scale in $eeqq$ contact interactions,
    the ratio of leptoquark
    mass to the Yukawa coupling for heavy leptoquark models and
    the mass scale parameter in models with 
    large extra dimensions. 
    The limit on the quark charge radius, in the classical form 
    factor approximation, is $0.85\cdot 10^{-16} \cm $.
     }

\makezeustitle

%% file: DESY-03-218-aut.tex
\def\3{\ss}                                                                                        
\pagenumbering{Roman}                                                                              
                                                   %
\begin{center}                                                                                     
{                      \Large  The ZEUS Collaboration              }                               
\end{center}                                                                                       
  S.~Chekanov,                                                                                     
  M.~Derrick,                                                                                      
  D.~Krakauer,                                                                                     
  J.H.~Loizides$^{   1}$,                                                                          
  S.~Magill,                                                                                       
  S.~Miglioranzi$^{   1}$,                                                                         
  B.~Musgrave,                                                                                     
  J.~Repond,                                                                                       
  R.~Yoshida\\                                                                                     
 {\it Argonne National Laboratory, Argonne, Illinois 60439-4815}, USA~$^{n}$                       
\par \filbreak                                                                                     
  M.C.K.~Mattingly \\                                                                              
 {\it Andrews University, Berrien Springs, Michigan 49104-0380}, USA
% ...
\newpage                                                                                           
$^{\    1}$ also affiliated with University College London, London, UK \\                          
% ...
                                                           %
                                                           %
\newpage   
                                                           %
                                                           %
\begin{tabular}[h]{rp{14cm}}                                                                       
$^{a}$ &  supported by the Natural Sciences and Engineering Research                               
          Council of Canada (NSERC) \\                                                             
$^{b}$ &  supported by the German Federal Ministry for Education and                               
          Research (BMBF), under contract numbers HZ1GUA 2, HZ1GUB 0, HZ1PDA 5, HZ1VFA 5\\         
$^{c}$ &  supported by the MINERVA Gesellschaft f\"ur Forschung GmbH, the                          
          Israel Science Foundation, the U.S.-Israel Binational Science                            
          Foundation and the Benozyio Center                                                       
          for High Energy Physics\\                                                                
$^{d}$ &  supported by the German-Israeli Foundation and the Israel Science                        
          Foundation\\                                                                             
$^{e}$ &  supported by the Italian National Institute for Nuclear Physics (INFN) \\                
$^{f}$ &  supported by the Japanese Ministry of Education, Culture,                                
          Sports, Science and Technology (MEXT) and its grants for                                 
          Scientific Research\\                                                                    
$^{g}$ &  supported by the Korean Ministry of Education and Korea Science                          
          and Engineering Foundation\\                                                             
$^{h}$ &  supported by the Netherlands Foundation for Research on Matter (FOM)\\                   
$^{i}$ &  supported by the Polish State Committee for Scientific Research,                         
          grant no. 620/E-77/SPB/DESY/P-03/DZ 117/2003-2005\\                                      
$^{j}$ &  partially supported by the German Federal Ministry for Education                         
          and Research (BMBF)\\                                                                    
$^{k}$ &  partly supported by the Russian Ministry of Industry, Science                            
          and Technology through its grant for Scientific Research on High                         
          Energy Physics\\                                                                         
$^{l}$ &  supported by the Spanish Ministry of Education and Science                               
          through funds provided by CICYT\\                                                        
$^{m}$ &  supported by the Particle Physics and Astronomy Research Council, UK\\                   
$^{n}$ &  supported by the US Department of Energy\\                                               
$^{o}$ &  supported by the US National Science Foundation\\                                        
$^{p}$ &  supported by the Polish State Committee for Scientific Research,                         
          grant no. 112/E-356/SPUB/DESY/P-03/DZ 116/2003-2005,2 P03B 13922\\                       
$^{q}$ &  supported by the Polish State Committee for Scientific Research,                         
          grant no. 115/E-343/SPUB-M/DESY/P-03/DZ 121/2001-2002, 2 P03B 07022\\                    
\end{tabular}                                                                                      
                                                           %
                                                           %

%% file: DESY-03-218-txt.tex
\pagenumbering{arabic} 
\pagestyle{plain}
%
%
\section{Introduction}
\label{sec:int}

 The HERA $ep$ collider has extended the kinematic range of deep
inelastic scattering (DIS) measurements by two orders of magnitude
in $Q^2$, the negative square of the four-momentum transfer,
compared to fixed-target experiments.
At values of $Q^2$ of about  $4\times10^4\gev^2$,
the $eq$ interaction, where $q$ is a 
constituent quark of the proton, is probed at distances
of $\sim 10^{-16}\cm$.
Measurements in this domain allow searches for new physics processes 
with characteristic mass scales in the $\tev$ range. 
New interactions between $e$ and $q$ involving mass scales 
above the center-of-mass energy can modify the cross section at high $Q^2$
via virtual effects, resulting in observable deviations from 
the Standard Model (SM) predictions.
Many such interactions, 
such as processes mediated by heavy leptoquarks,
can be modelled as four-fermion contact interactions. 
The SM predictions for $ep$ scattering 
in the  $Q^2$ domain of this study result from the evolution of 
accurate measurements of the proton structure functions made at
lower $Q^2$.
In this paper, a common method is applied to search for four-fermion 
interactions, for graviton exchange in models with large extra dimensions,
and for a finite charge radius of the quark.

In an analysis of 1994-97 $e^{+}p$ data  \cite{epj:c14:239}, 
the ZEUS Collaboration set limits on the effective mass scale for 
several parity-conserving compositeness models. 
Results presented here are based on approximately $130\pbi$ of
$e^{+}p$ and $e^{-}p$ data collected by ZEUS in the years 1994-2000.
Since this publication also includes the early ZEUS data, the results
presented here supersede those of the earlier publication \cite{epj:c14:239}.


\section{Standard Model cross section}
\label{sec:smxsec}

The differential SM cross section for  
neutral current (NC) $ep$ scattering, $e^\pm p \rightarrow e^\pm X$,
can be expressed in
terms of the kinematic variables $Q^2$, $x$ and $y$,
which are defined by the four-momenta 
of the incoming electron\footnote{Unless otherwise specified,
`electron' refers to both positron and electron.} ($k$), 
the incoming proton ($P$), and the scattered electron ($k'$) as
$Q^2=-q^2=-(k-k')^2$, $x=Q^2/(2q\cdot P)$, and $y=(q\cdot P)/(k\cdot P)$. 
For unpolarized beams, the leading-order electroweak 
cross sections can be expressed as
\begin{eqnarray}
  \frac{d^2\sigma^\NC(e^\pm p)}{dx\,dQ^2}(x,Q^2)&=&
        \frac{2\pi\alpha^2}{xQ^4}\,\left[\left(1+(1-y)^2\right)\,F_2^\NC\mp
        \left(1-(1-y)^2\right)\,xF_3^\NC\strut\right]\;,
                                                        \label{eq-lonc}
\end{eqnarray}
where $\alpha$ is the electromagnetic coupling constant.
The contribution of the longitudinal structure function, $F_L(x,Q^2)$, is
negligible at high $Q^{2}$ and is not taken into account 
in this analysis. 
At leading order (LO) in QCD, the structure functions
$F_2^\NC$ and $xF_3^\NC$ are given by 
\begin{eqnarray}
  F_2^\NC(x,Q^2)&=&\sum_{q=u,d,s,c,b} A_q(Q^2)\;
        \left[xq(x,Q^2)+x\qbar(x,Q^2)\right]\;,  \nonumber \\
  xF_3^\NC(x,Q^2)&=&\sum_{q=u,d,s,c,b} B_q(Q^2)\;
        \left[xq(x,Q^2)-x\qbar(x,Q^2)\right]\;, \nonumber 
\end{eqnarray}
where $q(x,Q^2)$ and $\qbar(x,Q^2)$ are the parton densities for
quarks and antiquarks.
The functions $A_q$ and $B_q$ are defined as 
\begin{equation}
  \begin{split}
  A_q(Q^2)&=\frac{1}{2}\left[
        (V^L_q)^2+(V^R_q)^2+(A^L_q)^2+(A^R_q)^2\right]\;,\\
  B_q(Q^2)&=\phantom{\frac{1}{2}\,}
        (V^L_q)(A^L_q)-(V^R_q)(A^R_q) \;,\\
  \end{split}
                                 \nonumber
\end{equation}
where the coefficient functions $V^{L,R}_q$ and $A^{L,R}_q$ are given by:
\begin{equation}
  \begin{split}
        V^i_q&=Q_q-(v_e\pm a_e)\,v_q\,\chi_Z \;,\\
        A^i_q&=\phantom{Q_q}-(v_e\pm a_e)\,a_q\,\chi_Z \;,\\
        v_f  &=T^\sthr_f-2\stws\,Q_f\;,\\
        a_f  &=T^\sthr_f\;,\\
        \chi_Z  &=\frac{1}{4\stws\ctws}\frac{Q^2}{Q^2+M_Z^2}\;.
  \end{split}
                                                        \label{eq-fco}
\end{equation}
In Eq.~\eq{fco}, the superscript $i$ denotes the left ($L$) or right ($R$)
helicity projection of the lepton field; the plus (minus) sign in the
definitions of $V^i_q$ and $A^i_q$ is appropriate for $i=L$($R$). 
The coefficients $v_f$
and $a_f$ are the SM vector and axial-vector coupling constants of an electron
($f=e$) or quark ($f=q$); $Q_f$ and $T^\sthr_f$ denote the fermion charge and
third component of the weak isospin; $M_Z$ and $\tw$ are the mass 
of the $Z^0$ and the electroweak mixing angle, respectively.

%
%

\section{Models for new physics}
\label{sec:models}


\subsection{General contact interactions}
\label{sec:ci}

Four-fermion contact interactions (CI) represent an effective theory, 
which describes low-energy effects  due to physics at much higher 
energy scales. 
Such models would describe the effects of heavy leptoquarks,
additional heavy weak bosons, and electron or quark compositeness.
The CI approach is not renormalizable 
and is only valid in the low-energy limit.
As strong limits have already been placed on scalar and tensor 
contact interactions \cite{hab-91-01}, 
only vector currents are considered here.
They can be represented 
by additional terms in the Standard Model Lagrangian, viz:
\begin{eqnarray}
{\cal L}_{CI} & = & 
   \sum_{^{i,j=L,R}_{q=u,d,s,c,b}} 
   \eta^{eq}_{ij} (\bar{e}_{i} \gamma^{\mu} e_{i} )
                 (\bar{q}_{j} \gamma_{\mu} q_{j}) \; ,
	\label{eq-cilagr}
\end{eqnarray}
where the sum runs over electron and quark helicities
and quark flavors.
The couplings  $\eta^{eq}_{ij}$ describe 
the helicity and flavor structure of contact interactions.
The CI Lagrangian (Eq.~\eq{cilagr}) results in the following modification of
the functions $V^i_q$ and $A^i_q$ of Eq.~\eq{fco}:
\begin{equation}
  \begin{split}
        V^i_q&=Q_q-(v_e\pm a_e)\,v_q\,\chi_Z+
                    \frac{Q^2}{2\alpha}(\eta_{iL}^{eq}+\eta_{iR}^{eq})\;,\\
        A^i_q&=\phantom{Q_q}-(v_e\pm a_e)\,a_q\,\chi_Z+
                    \frac{Q^2}{2\alpha}(\eta_{iL}^{eq}-\eta_{iR}^{eq})\;.\\
  \end{split}
                                     \nonumber
\end{equation}

It was assumed that all up-type quarks have the same
contact-interaction couplings, and a similar assumption was made 
for down-type quarks\footnote{The results depend very 
weakly on this assumption since heavy quarks make only a very small
contribution to high-$Q^{2}$ cross sections.
In most cases, the same mass-scale limits were obtained for CI scenarios
where only first-generation quarks are considered.
The largest difference between the obtained mass-scale limits is about 2\%.}:
\begin{eqnarray}
\eta^{eu}_{ij} & = \eta^{ec}_{ij} & = \eta^{et}_{ij} \;, \nonumber \\ 
\eta^{ed}_{ij} & = \eta^{es}_{ij} & = \eta^{eb}_{ij} \;, \nonumber
\end{eqnarray}
leading to eight independent couplings, $\eta^{eq}_{ij}$, with $q=u,d$. 
Due to the impracticality of setting limits
in an eight-dimensional space, a set of representative scenarios was
analyzed. 
Each scenario is defined by a set of eight coefficients,
$\epsilon^{eq}_{ij}$, each of which may take the values $\pm1$ or
zero, and the compositeness scale $\Lambda$.
The couplings are then defined by
\begin{eqnarray}
\eta^{eq}_{ij} & = & \epsilon^{eq}_{ij} \;
                   \frac{4 \pi}{\Lambda^{2}} \; . \nonumber
\end{eqnarray}
Note that models that differ in the overall sign of the
coefficients $\epsilon^{eq}_{ij}$ are distinct because of the
interference with the SM.


In this paper, different chiral structures of CI are  considered,
as listed in \tab{ci}.
Models listed in the lower part of the table were previously considered
in  the  published analysis of 1994-97 $e^{+}p$ data \cite{epj:c14:239}.
They fulfill the relation
\begin{eqnarray}
\eta^{eq}_{LL} + \eta^{eq}_{LR} - \eta^{eq}_{RL} - \eta^{eq}_{RR}
                    & = &  0 \; , \nonumber
\end{eqnarray}
which was imposed to conserve parity, and thereby complement strong 
limits from atomic parity violation (APV) results 
\cite{science:275:1759,pr:d45:1602}.
Since a later APV analysis \cite{prl:82:2484} indicated possible 
deviations from SM predictions,
models that violate parity,  listed in the upper part of \tab{ci},
have also been incorporated in the analysis.
The reported 2.3$\sigma$ deviation \cite{prl:82:2484} from the SM 
was later reduced to around 1$\sigma$,  after reevaluation of some 
of the theoretical corrections \cite{prl:85:1618,prl:86:3260}.


\subsection{Leptoquarks}
\label{sec:lq}

Leptoquarks (LQ) appear in certain extensions of the SM that 
connect leptons and quarks; they carry both lepton and 
baryon numbers and have spin 0 or 1. 
According to the general classification 
proposed by Buchm\"uller, R\"uckl and Wyler \cite{pl:b191:442},
there are 14 possible LQ states: seven scalar and seven
vector\footnote{Leptoquark states are named according to 
the so-called Aachen notation \cite{zfp:c46:679}.}.
In the limit of heavy LQs ($M_{LQ} \gg \sqrt{s}$),
the effect of $s$- and $t$-channel LQ exchange
is equivalent to a vector-type $eeqq$ contact interaction\footnote{
%
%
For the invariant mass range accessible at HERA, $\sqrt{s}\sim 300\gev$,
heavy LQ approximation is applicable for $M_{LQ} > 400\gev$.
For ZEUS limits covering LQ masses below $400\gev$ see \cite{pr:d68:052004}.
}.
The effective contact-interaction couplings, $\eta^{eq}_{ij}$, are
proportional to the square of the ratio of the leptoquark Yukawa 
coupling, $\lambda_{LQ}$, to the leptoquark mass, $M_{LQ}$:
\begin{eqnarray}
\eta^{eq}_{ij} & = & 
   a^{eq}_{ij}\left(\frac{\lambda_{LQ}}{M_{LQ}} \right)^{2} \; , \nonumber
\end{eqnarray}
where the coefficients $a^{eq}_{ij}$ depend on the LQ species
\cite{zfp:c74:595} and are twice as large for vector as for 
scalar leptoquarks. 
Only first-generation leptoquarks are considered in this analysis,
$q = u,d$.
The coupling structure for different leptoquark species is
shown in \tab{lq}.
Leptoquark models $S_0^L$ and $\tilde{S}^L_{1/2}$ correspond to
the squark states $\tilde{d_R}$ and $\overline{\tilde{u}_{L}}$,
in minimal supersymmetric theories with broken R-parity.


\subsection{Large extra dimensions}
\label{sec:led}

Arkani-Hamed, Dimopoulos and Dvali 
\cite{pl:b429:263,pl:b436:257,pr:d59:086004}
have proposed a model to
solve the hierarchy problem, 
assuming that space-time has $4+n$ dimensions. 
Particles, including strong and
electroweak bosons, are confined to four dimensions, but gravity can
propagate into the extra dimensions.
The extra $n$ spatial dimensions are compactified with a radius $R$.
The Planck scale, $M_P \sim 10^{19}\gev$, in $4$ dimensions 
is an effective scale arising from 
the fundamental Planck scale $M_D$ in $D=4+n$ dimensions.
The two scales are related by:
\begin{eqnarray}
	M_P^2 & \sim & R^n M_D^{2+n}\;. \nonumber
\end{eqnarray}
For extra dimensions with $R \sim 1\mm$ for $n=2$, 
the  scale $M_{D}$ can be of the order of $\tev$. 
At high energies, the strengths of the gravitational and
electroweak interactions can then become comparable.
After summing the effects of graviton excitations 
in the extra dimensions,  the graviton-exchange contribution 
to $eq \rightarrow eq$ scattering can be described as a 
contact interaction with an effective coupling 
strength of\cite{np:b544:3,pl:b460:383}
\begin{eqnarray}
	\eta_{G} & = & \frac{\lambda}{M_{S}^4}\; , \nonumber
\end{eqnarray}
where $M_S$ is an ultraviolet cutoff scale, 
expected to be of the order of $M_D$,
and the coupling $\lambda$ is of order unity.
Since the sign of $\lambda$ is not known {\it a priori}, both
values $\lambda = \pm 1$ are considered in this analysis.
However, due to additional energy-scale dependence, 
reflecting the number of accessible graviton excitations,
these contact interactions are not equivalent to the vector 
contact interactions of Eq. \eq{cilagr}.
To describe the effects of graviton exchange,
terms arising from pure graviton exchange ($G$),  
graviton-photon interference ($\gamma G$) and graviton-$Z$
($ZG$) interference
have to be added  to the SM $eq \rightarrow eq$ 
scattering cross section  \cite{pl:b479:358}:
\begin{eqnarray*}
  \frac{d \sigma(e^{\pm} q \rightarrow e^\pm q)}{d \hat{t}} & = & 
              \frac{d \sigma^{SM}}{d\hat{t}} + \frac{d \sigma^{G}}{d\hat{t}}
 + \frac{d \sigma^{\gamma G}}{d\hat{t}}  + \frac{d \sigma^{ZG}}{d\hat{t}}, \\
   \frac{d \sigma^{G}}{d\hat{t}} & = & 
    \frac{\pi \lambda^2}{32 M_S^8} \frac{1}{\hat{s}^2} 
     \left\{ 32\hat{u}^4 + 64\hat{u}^3\hat{t} + 42\hat{u}^2\hat{t}^2 + 
                             10\hat{u}\hat{t}^3+ \hat{t}^4 \right\}, \\   
  \frac{d \sigma^{\gamma G}}{d\hat{t}} & = & 
        \mp \frac{\pi \lambda}{2 M_S^4} \frac{\alpha Q_q}{\hat{s}^2} 
                        \frac{(2\hat{u} + \hat{t})^3}{\hat{t}},  \\     
   \frac{d \sigma^{Z G}}{d\hat{t}} & = & 
   \frac{\pi \lambda}{2 M_S^4} \frac{\alpha}{\hat{s}^2 \sin^2{2\theta_W}}
\left\{ \pm v_e v_q \frac{(2\hat{u}+\hat{t})^3}{\hat{t} - M^2_Z} 
   - a_e a_q \frac{\hat{t} (6\hat{u}^2 + 6\hat{u}\hat{t} 
               + \hat{t}^2)}{\hat{t}-M^2_Z}  \right\},
\end{eqnarray*}
where $\hat{s}$, $\hat{t}$ and $\hat{u}$, with $\hat{t} = -Q^2$, 
are the Mandelstam variables, 
while the other coefficients are given in Eq.~\eq{fco}.
The corresponding cross sections for $e^{\pm}\bar{q}$ scattering are obtained
by changing the sign of $Q_{q}$ and $v_{q}$ parameters.

Graviton exchange also
contributes to  electron-gluon scattering, $eg \rightarrow eg$, which is
not present at leading order in the SM:
\begin{eqnarray*}
   \frac{d \sigma(e^\pm g \rightarrow e^\pm g)}{d \hat{t}} & = & 
      \frac{\pi \lambda^2}{2 M_S^8} \frac{\hat{u}}{\hat{s}^2} 
      \left\{ 2\hat{u}^3 + 4\hat{u}^2\hat{t} + 3\hat{u}\hat{t}^2 
                                                 +\hat{t}^3 \right\} \, .
\end{eqnarray*}

 For a given point in the
$(x,Q^2)$ plane, the $e^{\pm}p$ cross section is then given by
\begin{eqnarray*}
\frac{d^2 \sigma(e^{\pm}p \to e^{\pm}X)}{d x d Q^2} (x, Q^2) & = &
       q(x,Q^2)\frac{d\sigma(e^{\pm}q)}{d \hat{t}} + 
 \bar{q}(x,Q^2)\frac{d\sigma(e^{\pm}\bar{q})}{d \hat{t}} +
       g(x,Q^2)\frac{d\sigma(e^{\pm}g)}{d \hat{t}}\; ,
\end{eqnarray*}
where $q(x,Q^2)$, $\bar{q}(x,Q^2)$ and $g(x,Q^2)$ 
are the quark, anti-quark and 
gluon densities in the proton, respectively.


\subsection{Quark form factor}
\label{sec:rq}

Quark substructure can be detected by  measuring
the spatial distribution of the quark charge.
If $Q^{2} \ll 1/R_{e}^{2}$ and $Q^{2} \ll 1/R_{q}^{2}$,
the SM predictions for the cross sections are modified,
approximately, to:
\begin{eqnarray*}
\frac{d\sigma}{dQ^{2}} & = & 
\frac{d\sigma^{SM}}{dQ^{2}} \;
\left( 1 - \frac{R_{e}^{2}}{6} \, Q^{2} \right)^{2} \;
\left( 1 - \frac{R_{q}^{2}}{6} \, Q^{2} \right)^{2} \; ,
\end{eqnarray*}
where $R_{e}$ and $R_{q}$ are the root-mean-square radii 
of the electroweak charge of the electron and the quark, respectively.
%


\section{Data samples}
\label{sec:data}

The data used in this analysis were collected with the ZEUS detector 
at HERA and correspond to an integrated luminosity of $48\pbi$ and $63\pbi$ 
for  $e^{+}p$  collisions collected in 1994-97 and 1999-2000 respectively, 
and $16\pbi$  for  $e^{-}p$  collisions collected in 1998-99.
The 1994-97 data set was collected at $\sqrt{s}=300\gev$ 
and the 1998-2000 data sets were taken with $\sqrt{s}=318\gev$.

The analysis is based upon the final event samples used in previously
published cross section measurements~\cite{epj:c11:427,epj:c28:175,zeus:xsec}.
Only events with $Q^2 > 1000\gev^2$ are considered.
The SM predictions were taken from the simulated event
samples used in the cross section measurements, where selection cuts and
event reconstruction are identical to those applied to the data.
Neutral current DIS events were simulated using 
the {\sc Heracles}~\cite{cpc:69:155} 
program with {\sc Djangoh}~\cite{cpc:81:381,spi:www:djangoh11}
for electroweak radiative corrections and higher-order matrix elements,
and the color-dipole model of {\sc Ariadne}~\cite{cpc:71:15}  
for the QCD cascade and hadronization.
The ZEUS detector was simulated using a program based on 
{\sc Geant} 3.13 \cite{tech:cern-dd-ee-84-1}.
The details of the data selection and reconstruction,
and the simulation used can be found 
elsewhere~\cite{epj:c11:427,epj:c28:175,zeus:xsec}.  

The distributions of NC DIS events in $Q^{2}$,
measured separately for each of the three data sets,
are in good agreement with SM predictions calculated using the CTEQ5D
parameterization \cite{epj:c12:375, pr:d55:1280}
of the parton distribution functions (PDFs) of the proton.
The CTEQ5D parameterization is based on a global QCD analysis of the data on 
high energy lepton-hadron and hadron-hadron interactions, including
high-$Q^{2}$ H1 and ZEUS results based on the 1994 $e^{+}p$ data.
The ZEUS data used in the CTEQ analysis amount to less than 3\% 
of the sample considered in this analysis.
In general, SM predictions in the $Q^2$ range considered here
are dominantly determined by fixed-target data at 
$Q^2<100\gev^2$ and $x>0.01$ \cite{pr:d67:012007}.


\section{Analysis method}
\label{sec:method}


\subsection{Monte Carlo reweighting}
\label{sec:mcrew}

The contact interactions analysis 
was based on a comparison of the measured  $Q^{2}$ 
distributions with the predictions of the MC simulation.
The effects of each CI scenario are taken into account by 
reweighting each MC event of the type $ep \to eX$ with the 
weight
\begin{equation}
  w=\left.{
          \frac{
               \frac{d^2\sigma}{dx\,dQ^2}(\text{SM$+$CI})
               }{
               \frac{d^2\sigma}{dx\,dQ^2}(\text{SM})
               }
               }\right|_{{\rm true\ } x, Q^2}\;.
                                                        \label{eq-wdef}
\end{equation}
The weight $w$ was calculated as the ratio of 
the leading-order\footnote{Note that CIs constitute 
a non-renormalizable effective theory for which higher orders 
are not well defined. 
} 
cross sections,  Eq.~\eq{lonc},
evaluated at the true values of $x$ and $Q^2$ 
as determined from the four-momenta of the exchanged boson and 
the incident particles. 
In simulated events where a photon with energy $E_\gamma$ is radiated by
the incoming electron (initial-state radiation), the
electron energy is reduced by $E_\gamma$.
This approach guarantees that possible differences between the SM and the
CI model in event-selection efficiency and migration corrections are properly
taken into account.
Under the assumption that
the difference between the SM predictions and 
those of the model including contact interactions is small,
higher-order QCD and electroweak corrections, including radiative
corrections, are also accounted for.


\subsection{Limit-setting procedure}
\label{sec:limit}

  For each of the models of new physics described above,
it is possible to characterize the strength of the interaction by
a single parameter: 
$4\pi/\Lambda^2$ for contact interactions;
$(\lambda_{LQ}/M_{LQ})^2$ for leptoquarks;
$\lambda/M_S^4$ for models with large extra dimensions; 
and $R_q^2$ for the quark form factor. 
In the following, this parameter is denoted by $\eta$.
For contact interactions, models with large extra dimensions
and the quark form factor model,
scenarios with positive and negative $\eta$ values were
considered separately.

For a given model, the likelihood 
was calculated as
\begin{eqnarray}
       L(\eta) & = & \prod_{i} 
 e^{-\mu_i(\eta)} \cdot \frac{\mu_i(\eta)^{n_i}}{n_i !} \;,
                                                        \nonumber
\end{eqnarray}
where the product runs over all $Q^2$ bins, 
$n_i$ is the number of events observed in $Q^2$ bin $i$ and 
$\mu_i(\eta)$ is the expected number of events in that bin 
for a coupling strength $\eta$.
The likelihood for the complete $e^\pm p$ data set
was obtained by multiplying the likelihoods 
for each of the three running periods.

The value of $\eta$ for which $L(\eta)$ is maximized 
is denoted as $\eta_\circ$.
First $\eta^{\rm data}_{\circ}$, the value of
$\eta$ that best describes the observed $Q^2$ spectra
was determined. 
Using ensembles of Monte Carlo experiments (MCE),
the expected distribution of $\eta_\circ$
was then determined as a function of $\eta_{MC}$, the coupling value
used as the input to the simulation.
The \CL{95} limit on $\eta$ was defined as the value of
$\eta_{MC}$ for which
the probability that $|\eta_\circ| > |\eta_\circ^{\rm data}|$ was 0.95.

For each value of $\eta_{MC}$, the nominal number of events expected 
in each $Q^2$ bin $i$, denoted ${\tilde\mu}_i(\eta_{MC})$ was
calculated by reweighting the SM MC prediction according to Eq.~\eq{wdef}.
Theoretical and experimental systematic uncertainties
were taken into account by treating each uncertain quantity
as a random variable.
For each uncertainty, 100\% correlation between 
systematic variations in different bins was assumed.
For each individual MCE, an independent random variable,
$\delta_j$,  with zero mean,
was generated for each systematic uncertainty $j$. 
The expected number of events in each $Q^2$ bin $i$ was then given by
the product of the nominal expectation, ${\tilde\mu}_i$, and $N_{\rm sys}$
random factors which account for the uncertainties in the estimation
of $\mu_i$ as follows:
\begin{eqnarray*}
\mu_i & = & {\tilde\mu}_i(\eta_{MC}) \cdot \prod_{j=1}^{N_{\rm sys}}
\left( 1+c_{ij}\right)^{\delta_j}.
\end{eqnarray*}
The coefficent $c_{ij}$ is the fractional change in the expected
number of events in bin $i$ for a unit change in $\delta_j$.
This definition of $\mu_i$ reduces to a linear dependence of 
$\mu_i$ on each $\delta_j$ when $\delta_j$ is small, while avoiding
the possibility of $\mu_i$ becoming negative which would arise if
$\mu_i$ was defined as a linear function of the $\delta_j$'s.
For most of the systematic uncertainties, $\delta_j$ follows a Gaussian
distribution, 
except for a few where it follows a uniform distribution,
as noted in the next section.
For a Gaussian $\delta_j$ distribution,
the definition of $\mu_i$ corresponds to a Gaussian distribution 
in $\log \mu_i$.
About one million MCEs were generated for each  model,
so that the statistical error was negligible.


\subsection{Systematic uncertainties}

Uncertainties in the SM cross sections considered in this study
were estimated using the {\sc Epdflib} program \cite{epdflib}
based on {\sc Qcdnum} \cite{epj:c14:285}.
Fractional variations estimated from {\sc Epdflib} were used to rescale
the nominal SM expectations calculated with CTEQ5D.
The following uncertainties were included:
\begin{itemize}
  \item statistical and systematic uncertainties of the data 
        used as an input  to the NLO QCD fit.
        These errors were the largest uncertainty in the SM expectations.
        At high $Q^{2}$, the uncertainty 
 is up to about 4.5\% (3\%) for $e^{+}p$ ($e^{-}p$) data;

  \item uncertainty in the value of  $\alpha_{S} (M_{Z}^{2})$
        used in the NLO QCD fit.
        The resulting uncertainties of NC DIS cross sections 
        at high $Q^{2}$, estimated assuming an
        error on $\alpha_{S} (M_{Z}^{2})$ of  $\pm$0.002 \cite{pr:d66:010001},
        is about 1.6\%;

  \item uncertainties in the nuclear corrections applied to the deuteron
        data ($K_{D}$) and to the data from neutrino scattering on iron
       ($K_{Fe}$) used in {\sc Qcdnum}. As suggested in {\sc Epdflib}, 
 variations by up to 100\% for $K_{D}$ and 50\% for $K_{Fe}$ were applied,
        treating the corrections as uniformly distributed random
        variables.
        The corresponding uncertainties of NC DIS cross sections 
        at high $Q^{2}$, are up to about 
        1.7\% (0.8\%) for $K_{D}$ and up to about 
        3\% (0.7\%) for $K_{Fe}$, for $e^{+}p$ ($e^{-}p$) data.

\end{itemize}
The PDF uncertainties calculated using {\sc Epdflib} are
similar to those obtained from a ZEUS NLO QCD fit \cite{pr:d67:012007}, 
when high-$Q^2$ HERA data were excluded from the fit.

In addition to the uncertainty in the SM prediction, the following
experimental uncertainties were taken into account:
\begin{itemize}
  
  \item the scale uncertainty on the energy of the scattered electron
 of $\pm$(1--3)\% depending on the topology of the event  \cite{epj:c21:443}.
The resulting uncertainty of NC DIS cross section 
at high $Q^{2}$ is about 0.6\% (1.3\%), for $e^{+}p$ ($e^{-}p$) data;
  
 \item the uncertainty in the hadronic energy scale
of $\pm$(1--2)\% depending on the topology of the event \cite{pl:b539:197}.
The resulting cross section uncertainty 
at high $Q^{2}$ is about 1\%, for both $e^{+}p$ and $e^{-}p$ data;

  \item uncertainties on the luminosity measurement of
        1.6\% for the 1994-97 $e^{+}p$ data,
        1.8\% for the 1998-99 $e^{-}p$ data and
        2.5\% for the 1999-2000 $e^{+}p$ data.
    Correlations between luminosity uncertainties for different data-taking
    periods are small and were neglected in the analysis. 

\end{itemize}
As the double-angle method used to reconstruct the kinematics 
of the events~\cite{epj:c11:427,epj:c28:175,zeus:xsec}
is relatively insensitive to uncertainties in the absolute energy
scale of the calorimeter, 
the largest experimental uncertainty 
in the numbers of NC DIS events expected at high $Q^2$
is due to the luminosity measurement.


\section{Results}
\label{sec:res}

No significant deviation of the ZEUS data from the SM prediction using the 
CTEQ5D parameterization of the proton PDF was observed. 
For all models considered, the best description of the data was obtained
for very small values of $|\eta_\circ^{\rm data}|$, i.e. close to the SM.
The probability of obtaining larger best-fit coupling from the SM, 
i.e. the probability  that 
an experiment would produce a value of $|\eta_\circ|$ greater than 
that obtained from the data, $|\eta_\circ| > |\eta_\circ^{\rm data}|$, 
calculated with MCEs assuming the SM cross section,
was above 25\% in all cases.
Therefore, limits on the strength parameters of the models
described in Sec.~\ref{sec:models} are presented in this paper.


The measured $Q^{2}$ spectra for $e^{+}p$ and $e^{-}p$ data, 
normalized to the SM predictions are shown in Fig.~\ref{fig-vvaa}.
Also shown are curves, for VV and AA contact-interaction models
(Section~\ref{sec:ci}),
which correspond to the \CL{95} exclusion limits
on $\Lambda$.
The \CL{95} limits on the compositeness scale $\Lambda$, 
for different CI models, are compared in \fig{bars} and \tab{ci}.
Limits range from $1.7 \tev$ for the LL model 
to $6.2\tev$ for the VV model.  
Also indicated in the figure are the best-fit coupling values,
$\eta_\circ^{\rm data} = \frac{4 \pi}{\Lambda^2}$, 
for positive and negative couplings.
For comparison, the positions of the global likelihood maxima
with  $\pm 1 \sigma$ and $\pm 2 \sigma$ 
error\footnote{
Errors are calculated from the likelihood variation:
$\pm 1 \sigma$ and $\pm 2 \sigma$ errors 
correspond to the decrease of the likelihood value to 
$\log L(\eta) =  \log L(\eta_\circ) - \frac{1}{2}  $ 
and $\log L(\eta) =  \ \log L(\eta_\circ) - 2 $, respectively.} 
bars are included in \fig{bars}.
Systematic uncertainties are taken into account by
averaging the likelihood values over systematic uncertainties.
For most models, the $\pm 2 \sigma$ error bars are in good agreement with
\CL{95} limits calculated with the MCE approach.

The \CL{95} lower limits on the compositeness scale $\Lambda$  
are compared in \tab{ci}  with limits from the 
H1 collaboration\cite{pl:b568:35}, 
the Tevatron~\cite{prl:79:2198, prl:82:4769} and the
LEP~\cite{epj:c12:183, epj:c11:383, pl:b489:81, epj:c13:553} 
experiments (where only the results from $e^+e^- \to q\bar{q}$ 
channel are quoted). 
In \tab{ci} the relations between CI couplings 
for the compositeness models considered 
are also included. 
The results on the compositeness scale $\Lambda$ presented here are 
comparable to those obtained by 
other experiments, where they exist.
For many models, this analysis sets the only existing limits.


The leptoquark analysis takes into account LQs that couple
to the electron and the first-generation quarks ($u$, $d$) 
only (Section~\ref{sec:lq}).
Deviations in the $Q^{2}$ distribution of $e^{+}p$ and
$e^{-}p$ NC DIS events, corresponding to the \CL{95} exclusion limits
for selected scalar and vector leptoquark models,
are compared with ZEUS data in Fig.~\ref{fig-lqsv}.
The  \CL{95} limits on the ratio of the leptoquark mass to the Yukawa
coupling, $M_{LQ}/\lambda_{LQ}$, are summarized in \tab{lq} together with
the coefficients $a^{eq}_{ij}$ describing the CI coupling structure.
The  limits range from 0.27$\tev$ for $\tilde{S}^R_{\circ}$ model to
1.23$\tev$ for  $V^L_{1}$ model.
\tab{lq} also shows the LQ limits obtained by the
H1 collaboration~\cite{pl:b568:35} and by the 
LEP experiments~\cite{epj:c12:183, pl:b489:81}.
In general, comparable limits are obtained.
For the  $S_1^L$, $V^R_{1/2}$ and $\tilde{V}^L_{1/2}$
leptoquarks, the ZEUS  analysis provides the most stringent
limits.

When only the NC DIS event sample is considered,
the leptoquark limits obtained in the contact-interaction
approximation are similar to, or better than, the high-mass limits from
the ZEUS resonance-search analysis \cite{pr:d68:052004}.
However, for $S_0^L$, $S_1^L$ and $V_0^L$ models these previously 
published limits are more stringent, as the possible leptoquark 
contribution to charged current DIS was also taken into account.


For the model with large extra dimensions (Section~\ref{sec:led}),  
\CL{95} lower limits on the mass scale in $n$ dimensions of
\begin{eqnarray*}
     M_{S} & > & 0.78\tev \hspace*{3ex} {\rm for~} \lambda = +1\; , \\
     M_{S} & > & 0.79\tev \hspace*{3ex} {\rm for~} \lambda = -1\; ,
\end{eqnarray*}
were obtained.
In \fig{edrq}, effects of graviton exchange on the $Q^{2}$ distribution,
corresponding to these limits,
are compared with ZEUS $e^{+}p$ (\fig{edrq}a) 
and $e^{-}p$ (\fig{edrq}b) data.
The limits on $M_S$ obtained in this analysis are similar
to those obtained by the H1 collaboration~\cite{pl:b568:35} 
and stronger than limits from $q\bar{q}$ production at LEP~\cite{pl:b470:281}.
However, if all final states are considered, 
the limits derived from $e^+e^-$ collisions exceed $1\tev$ \cite{pl:b470:281}.
Limits above $1\tev$ are also obtained in $p\bar{p}$ from the measurement 
of $e^-e^+$ and $\gamma \gamma$ production \cite{prl:86:1156}.


Assuming the electron to be point-like ($R_e = 0 $), 
the \CL{95} upper limit on the effective quark-charge 
radius (Section~\ref{sec:rq}) of
\begin{eqnarray*}
	R_q & < & 0.85\cdot 10^{-16} \cm 
\end{eqnarray*}  
was obtained.
The present result improves the limits set in $ep$ scattering by the 
H1 collaboration~\cite{pl:b568:35} ($R_q < 1.0\cdot 10^{-16} \cm$) and  
is similar to the limit set by the CDF collaboration in 
$p\bar{p}$ collisions using the Drell-Yan production of $e^+e^-$ and
$\mu^+\mu^-$ pairs~\cite{prl:79:2198} 
($R_q < 0.79\cdot 10^{-16}\cm$).\footnote{Limits
on the effective quark radius published by the CDF collaboration
\cite{prl:79:2198} were calculated assuming $R_q = R_e$.
For comparison with limits assuming $R_e = 0 $, the limit value  was 
scaled by a factor $\sqrt{2}$.}
The L3 collaboration has presented a stronger limit 
($R_q < 0.42\cdot 10^{-16}\cm$, assuming $R_e = 0 $), based on 
quark-pair production measurement at LEP2 \cite{pl:b489:81} and
assuming the same effective charge radius 
for all produced quark flavors. 

If the charge distribution in the quark changes 
sign as a function of the radius, negative values can also be
considered for $R_{q}^{2}$.
For such a model, the ZEUS \CL{95} upper limit on the effective quark-charge 
radius squared can be written as:
\begin{eqnarray*}
	-R_q^2 & < & (1.06 \cdot 10^{-16} \cm)^{2} \; .
\end{eqnarray*}  
Cross section deviations corresponding to the \CL{95} exclusion limits
for the effective radius, $R_{q}$, of the electroweak charge of the
quark are compared with the ZEUS data in \fig{edrq}c.


\section{Conclusions}

 A search for signatures of physics beyond the Standard Model 
has been performed with  the  $e^{+}p$ and $e^{-}p$ data
collected by the ZEUS Collaboration in the years 1994-2000, with
integrated luminosities of 112 and $16\pbi$, 
reaching  $Q^2$ values as high as  $4\times10^4\gev^2$.
No significant deviation from Standard Model predictions was
observed and \CL{95} limits were obtained for the relevant 
parameters of the models studied.
For the contact-interaction models, 
limits on the effective mass scale, $\Lambda$ 
(i.e. compositeness scale), ranging from 1.7 to 6.2$\tev$ have been
obtained.
Limits ranging from 0.27 to 1.23$\tev$ have been set
for  the ratio of the leptoquark mass to the Yukawa coupling,
$M_{LQ}/\lambda_{LQ}$, in the limit of large leptoquark masses, 
$M_{LQ} \gg \sqrt{s}$. 
Limits were derived on the mass scale parameter in models with 
large extra dimensions: for positive (negative) coupling signs, scales below
0.78$\tev$ (0.79$\tev$) are excluded.
A quark-charge radius larger than $0.85\cdot 10^{-16} \cm $ has
been excluded, using the classical form-factor
approximation.

The limits derived in this analysis are comparable to the 
limits obtained by the H1 collaboration and by the LEP and Tevatron
experiments. 
For many models 
the analysis presented here provides the most stringent limits to date.

%
%
\setcounter{secnumdepth}{0}
\section{Acknowledgements}
 
This measurement was made possible by the inventiveness and the diligent
efforts of the HERA machine group.
The strong support and encouragement of the DESY directorate has been
invaluable.
The design, construction, and installation of the ZEUS detector has been made
possible by the ingenuity and dedicated effort of many people 
who are not listed as authors. Their
contributions are acknowledged with great appreciation.

\vfill\eject

%% file: DESY-03-218-ref.tex
{
\def\bibname{\Large\bf References}
\def\refname{\Large\bf References}
\pagestyle{plain}
\ifzeusbst
  \bibliographystyle{./BiBTeX/bst/l4z_default}
\fi
\ifzdrftbst
  \bibliographystyle{./BiBTeX/bst/l4z_draft}
\fi
\ifzbstepj
  \bibliographystyle{./BiBTeX/bst/l4z_epj}
\fi
\ifzbstnp
  \bibliographystyle{./BiBTeX/bst/l4z_np}
\fi
\ifzbstpl
  \bibliographystyle{./BiBTeX/bst/l4z_pl}
\fi
{\raggedright
\bibliography{./BiBTeX/user/syn.bib,%
              ./BiBTeX/bib/l4z_articles.bib,%
              ./BiBTeX/bib/l4z_books.bib,%
              ./BiBTeX/bib/l4z_conferences.bib,%
              ./BiBTeX/bib/l4z_h1.bib,%
              ./BiBTeX/bib/l4z_misc.bib,%
              ./BiBTeX/bib/l4z_old.bib,%
              ./BiBTeX/bib/l4z_preprints.bib,%
              ./BiBTeX/bib/l4z_replaced.bib,%
              ./BiBTeX/bib/l4z_temporary.bib,%
              ./BiBTeX/bib/l4z_zeus.bib,%
              ./DESY-03-218-ref.bib%
}}
}
\vfill\eject

%% file: DESY-03-218-tab.tex

\begin{sidewaystable}[tbp]
  \begin{center}
   \begin{tabular}{|c@{~~~[}r@{,}r@{,}r@{,}r|cc||cc||cc|cc||cc|cc|cc|}
\hline
  \multicolumn{7}{|c||}{{\bf ZEUS 1994-2000 $e^\pm p$ \CL{95} (\tev)}} & 
  \multicolumn{2}{c||}{H1} & \multicolumn{2}{c|}{\DO} & 
  \multicolumn{2}{c||}{CDF} & \multicolumn{2}{c|}{ALEPH} & 
  \multicolumn{2}{c|}{L3} & \multicolumn{2}{c|}{OPAL} \\
\hline
 \multicolumn{5}{|r|}{Coupling structure}
 & &  & &  & & & & & & & & & & \\
  Model & 
 $\epsilon_{_{LL}}$ & $\epsilon_{_{LR}}$  &
                 $\epsilon_{_{RL}}$ &  $\epsilon_{_{RR}}$]~~ &
~~~$\Lambda^{-}$  &  $\Lambda^{+}$ & $\Lambda^{-}$  &  $\Lambda^{+}$ & 
   $\Lambda^{-}$  &  $\Lambda^{+}$ & $\Lambda^{-}$  &  $\Lambda^{+}$ & 
   $\Lambda^{-}$  &  $\Lambda^{+}$ & $\Lambda^{-}$  &  $\Lambda^{+}$ & 
   $\Lambda^{-}$  &  $\Lambda^{+}$ \\
\hline
LL 
 &  +1  &  0  &  0  &  0]~~      
 & 1.7 & 2.7 & 1.6 & 2.8 & 4.2 & 3.3 & 3.7 & 2.5 & 6.2 & 5.4 & 2.8 & 4.2 & 3.1 & 5.5  \\
LR 
 &  0  &  +1  &  0  &  0]~~      
 & 2.4 & 3.6 & 1.9 & 3.3 & 3.6 & 3.4 & 3.3 & 2.8 & 3.3 & 3.0 & 3.5 & 3.3 & 4.4 & 3.8  \\
RL 
 &  0  &  0  &  +1  &  0]~~      
 & 2.7 & 3.5 & 2.0 & 3.3 & 3.7 & 3.3 & 3.2 & 2.9 & 4.0 & 2.4 & 4.6 & 2.5 & 6.4 & 2.7  \\
RR 
 &  0  &  0  &  0  &  +1]~~      
 & 1.8 & 2.7 & 2.2 & 2.8 & 4.0 & 3.3 & 3.6 & 2.6 & 4.4 & 3.9 & 3.8 & 3.1 & 4.9 & 3.5  \\
\hline
VV 
 &  +1  &  +1  &  +1  &  +1]~~    
 & 6.2 & 5.4 & 5.5 & 5.3 & 6.1 & 4.9 & 5.2 & 3.5 & 7.1 & 6.4 & 5.5 & 4.2 & 7.2 & 4.7  \\
AA 
 &  +1  &  $-1$  &  $-1$  &  +1]~~ 
 & 4.7 & 4.4 & 4.1 & 2.5 & 5.5 & 4.7 & 4.8 & 3.8 & 7.9 & 7.2 & 3.8 & 6.1 & 4.2 & 8.1  \\
VA 
 &  +1  &  $-1$  &  +1  &  $-1$]~~ 
 & 3.3 & 3.2 & 3.0 & 2.9 & & & & & & & & & &  \\
X1 
 &   +1  &  $-1$  &  0  &  0]~~ 
 & 3.6 & 2.6 & & & 4.5 & 3.9 & & & & & & & &  \\
X2 
 &   +1  &  0  &  +1  &  0]~~ 
 & 3.9 & 4.0 & & & & & & & & & & & &  \\
X3 
 &   +1  &  0  &  0  &  +1]~~ 
 & 3.7 & 3.6 & 3.9 & 3.7 & 5.1 & 4.2 & & & 7.4 & 6.7 & 3.7 & 4.4 & 4.4 & 5.4  \\
X4 
 &   0  &  +1  &  +1  &  0]~~ 
 & 5.1 & 4.8 & 4.4 & 4.4 & 4.4 & 3.9 & & & 4.5 & 2.9 & 5.2 & 3.1 & 7.1 & 3.4  \\
X5 
 &   0  &  +1  &  0  &  +1]~~ 
 & 4.0 & 4.0 & & & & & & & & & & & &  \\
X6 
 &   0  &  0  &  +1  &  $-1$]~~ 
 & 2.5 & 3.5 & & & 4.3 & 4.0 & & & & & & & &  \\
U1 
 &   +1  &  $-1$  &  0  &  0]$^{eu}$ 
 & 3.8 & 3.6 & & & & & & & & & & & &  \\
U2 
 &   +1  &  0  &  +1  &  0]$^{eu}$ 
 & 5.0 & 4.2 & & & & & & & & & & & &  \\
U3 
 &   +1  &  0  &  0  &  +1]$^{eu}$ 
 & 5.0 & 4.1 & & & & & & & & & 5.2 & 9.2 & &  \\
U4 
 &   0  &  +1  &  +1  &  0]$^{eu}$ 
 & 5.8 & 4.8 & & & & & & & & & 3.2 & 2.3 & &  \\
U5 
 &   0  &  +1  &  0  &  +1]$^{eu}$ 
 & 5.2 & 4.3 & & & & & & & & & & & &  \\
U6 
 &   0  &  0  &  +1  &  $-1$]$^{eu}$ 
 & 2.8 & 3.4 & & & & & & & & & & & &  \\
\hline
    \end{tabular}
  \end{center}
  \caption{
     Coupling structure
  $[\epsilon_{LL},\epsilon_{LR}, \epsilon_{RL}, \epsilon_{RR}]$ of the 
         compositeness models and the \CL{95} limits on
         the compositeness scale, $\Lambda$, resulting from the ZEUS
       analysis of 1994-2000 $e^{\pm}p$ data. Each row of the table represents
         two scenarios corresponding to $\eta>0$ ($\Lambda^{+}$) and
         $\eta<0$ ($\Lambda^{-}$). The same coupling structure applies
        to d and u quarks, except for 
	the models U1 to U6, for which the couplings for the d quarks are zero.
        Also shown are results obtained by the H1 collaboration, 
the $p\bar{p}$ collider experiments \DO and CDF, and the LEP experiments
ALEPH, L3 and OPAL. For the LEP experiments, limits derived from the channel
	$e^+e^- \to q\bar{q}$ are quoted.
        }
  \label{tab-ci}
\end{sidewaystable}


\begin{table}[btp]
\begin{center}
\begin{tabular}{|cl|c||c|cc|}
\hline
\multicolumn{3}{|c||}{{\bf ZEUS 1994-2000 $e^\pm p$ \CL{95}}} & 
 \multicolumn{3}{c|}{$M_{LQ}/\lambda_{LQ}$ (\tev)}  \\
Model & 
Coupling Structure & 
$M_{LQ}/\lambda_{LQ}$ (\tev) & H1 & L3 & OPAL  \\
\hline
$S_0^L$ 
 &  $a^{eu}_{_{LL}}=+\frac{1}{2}$                     
 & 0.61 & 0.71 & 1.40 & 0.98  \\
$S_0^R$ 
 &  $a^{eu}_{_{RR}}=+\frac{1}{2}$                     
 & 0.56 & 0.64 & 0.30 & 0.30  \\
$\tilde{S}_0^{R}$ 
 &  $a^{ed}_{_{RR}}=+\frac{1}{2}$           
 & 0.27 & 0.33 & 0.58 & 0.80  \\
$S_{1/2}^L$ 
 &  $a^{eu}_{_{LR}}=-\frac{1}{2}$                       
 & 0.83 & 0.85 & 0.54 & 0.74  \\
$S_{1/2}^R$ 
 &  $a^{ed}_{_{RL}}=a^{eu}_{_{RL}}=-\frac{1}{2}$        
 & 0.53 & 0.37 &  & 0.86  \\
$\tilde{S}_{1/2}^{L}$ 
 &  $a^{ed}_{_{LR}}=-\frac{1}{2}$             
 & 0.43 & 0.43 & 0.42 & 0.48  \\
$S_{1}^{L}$ 
 &  $a^{ed}_{_{LL}}=+1, \; a^{eu}_{_{LL}}=+\frac{1}{2}$ 
 & 0.52 & 0.49 &  &   \\
\hline
$V_0^L$ 
 &  $a^{ed}_{_{LL}}=-1$                               
 & 0.55 & 0.73 & 1.83 & 1.27  \\
$V_0^R$ 
 &  $a^{ed}_{_{RR}}=-1$                               
 & 0.47 & 0.58 & 0.51 & 0.54  \\
$\tilde{V}_0^{R}$ 
 &  $a^{eu}_{_{RR}}=-1$                     
 & 0.87 & 0.99 & 1.02 & 1.44  \\
$V_{1/2}^L$ 
 &  $a^{ed}_{_{LR}}=+1$                                 
 & 0.47 & 0.42 & 0.71 & 0.90  \\
$V_{1/2}^R$ 
 &  $a^{ed}_{_{RL}}=a^{eu}_{_{RL}}=+1$                  
 & 0.99 & 0.95 &  & 0.71  \\
$\tilde{V}_{1/2}^{L}$ 
 &  $a^{eu}_{_{LR}}=+1$                       
 & 1.06 & 1.02 & 0.54 & 0.59  \\
$V_{1}^{L}$ 
 &  $a^{ed}_{_{LL}}=-1, \; a^{eu}_{_{LL}}=-2 $          
 & 1.23 & 1.36 &  &   \\
\hline
\end{tabular}
\end{center}
  \caption{Coefficients $a^{eq}_{ij}$ defining the effective 
           leptoquark couplings in the contact-interaction limit
           $M_{LQ}\gg \sqrt{s}$ and the \CL{95} lower limits on
       the leptoquark mass to the Yukawa coupling ratio $M_{LQ}/\lambda_{LQ}$
       resulting from the CI analysis of the ZEUS 1994-2000 $e^{\pm}p$ data,
        for different models of scalar (upper part of the table)
       and vector (lower part) leptoquarks.
     Also shown are results obtained by the H1 collaboration 
     and corresponding contact-interaction limits 
   from the LEP experiments L3 and OPAL.
  The limits from LEP on the compositeness scale $\Lambda$, for models with
 coupling structure corresponding to those of scalar (vector) leptoquarks, 
  were scaled by factor $1/\sqrt{8\pi}$ ($1/\sqrt{4\pi}$).
       }
  \label{tab-lq}
\end{table}

%% file: DESY-03-218-fig.tex

\begin{figure}[tbp]
\begin{center}
\epsfig{file=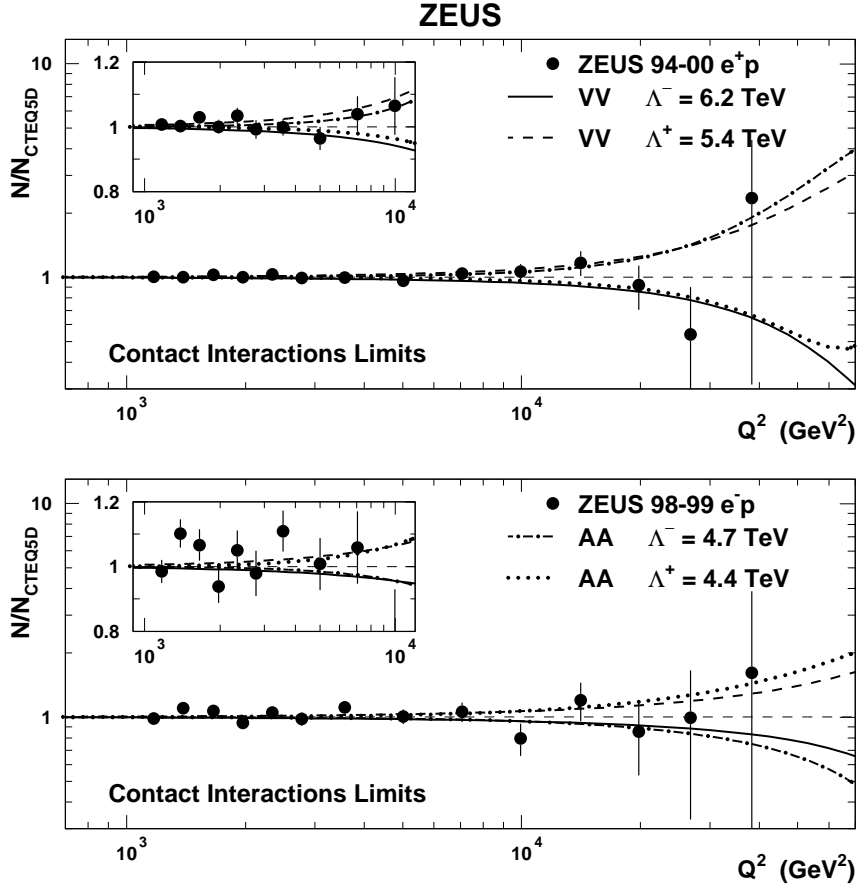,width=12cm}
\end{center}
  \caption{
          ZEUS data compared with \CL{95} exclusion limits for 
 the effective mass scale in the VV and AA contact-interaction models,
     for positive ($\Lambda^{+}$) and negative  ($\Lambda^{-}$)  couplings. 
  Results are normalized to the Standard Model expectations
  calculated using the CTEQ5D parton distributions.
  The insets show the comparison in the $Q^2 < 10^4 \gev^2$ region,
  with a linear ordinate scale.
         }
  \label{fig-vvaa}
\end{figure}


\begin{figure}[tbp]
\begin{center}
\epsfig{file=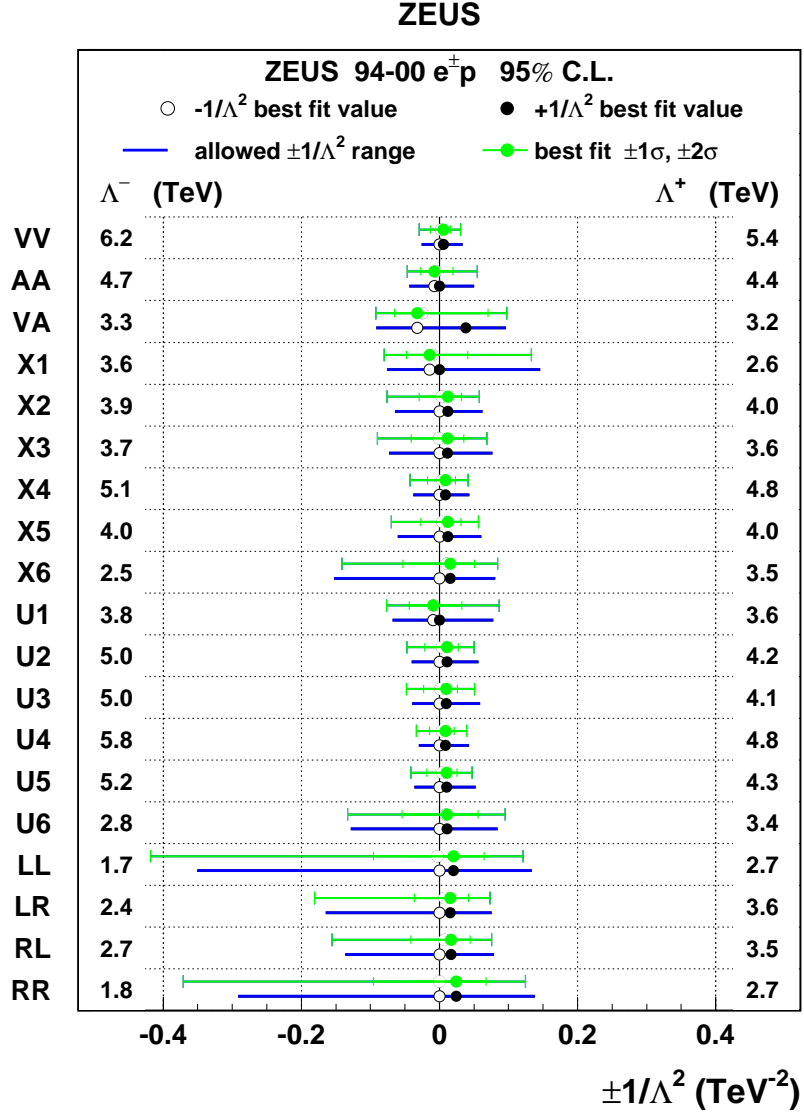,width=12cm}
\end{center}
  \caption{
          Confidence intervals of 
          $\pm 1/\Lambda^2$ at $95\%$ C.L.\ for general 
           CI scenarios studied in this paper (dark horizontal bars). 
           The numbers at the right (left) margin 
           are the corresponding lower limits on the
           mass scale $\Lambda^+$ ($\Lambda^-$).
           The dark filled (open) circles indicate 
           the positions corresponding to the best-fit coupling values, 
           $\eta_\circ^{\rm data}$, for positive (negative)  couplings. 
     The light filled circles with error bars indicate 
    the position of the global likelihood maximum. 
    For calculation of  $\pm 1 \sigma$ and $\pm 2 \sigma$ errors
    on the global maximum position, likelihood values are averaged over
   systematic uncertainties.
           }
  \label{fig-bars}
\end{figure}


\begin{figure}[tbp]
\begin{center}
\epsfig{file=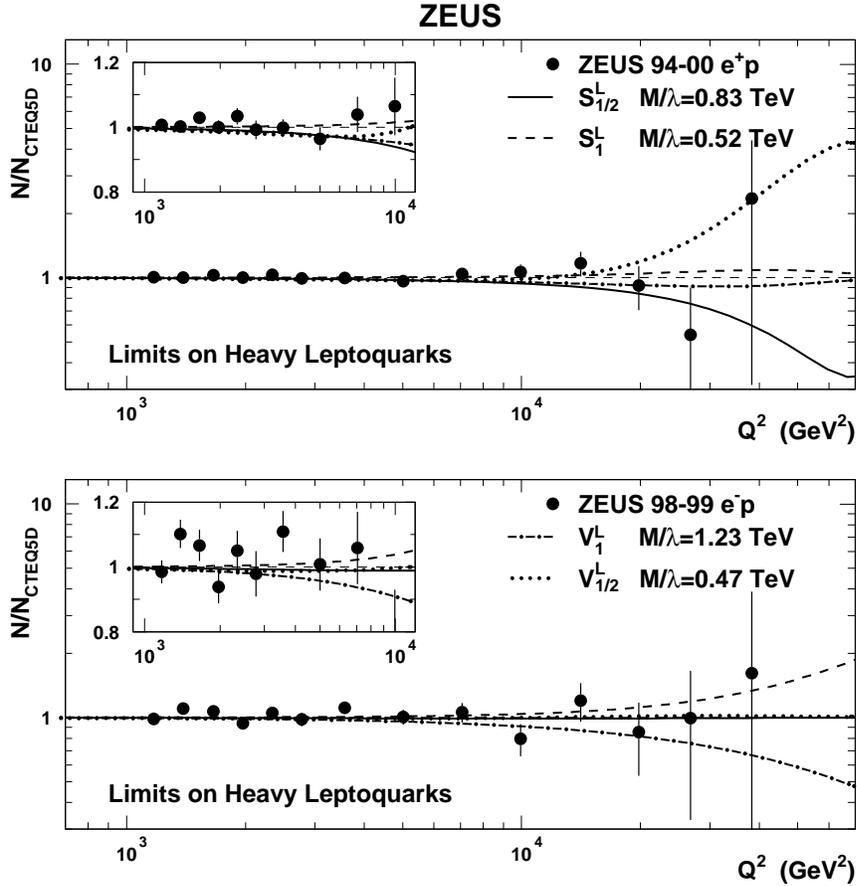,width=12cm}
\end{center}
  \caption{ 
          ZEUS data compared with \CL{95} exclusion limits for the
          ratio of the leptoquark mass to the Yukawa coupling,
          $M/\lambda$, for the $S^{L}_{1/2}$, $S_{1}^L$, $V_1^L$ 
          and $V^L_{1/2}$ leptoquarks. 
	  Results are normalized to the Standard Model expectations
	  calculated using the CTEQ5D parton distributions. 
  The insets show the comparison in the $Q^2 < 10^4 \gev^2$ region,
  with a linear ordinate scale.
         }
  \label{fig-lqsv}
\end{figure}


\begin{figure}[tbp]
\begin{center}
\epsfig{file=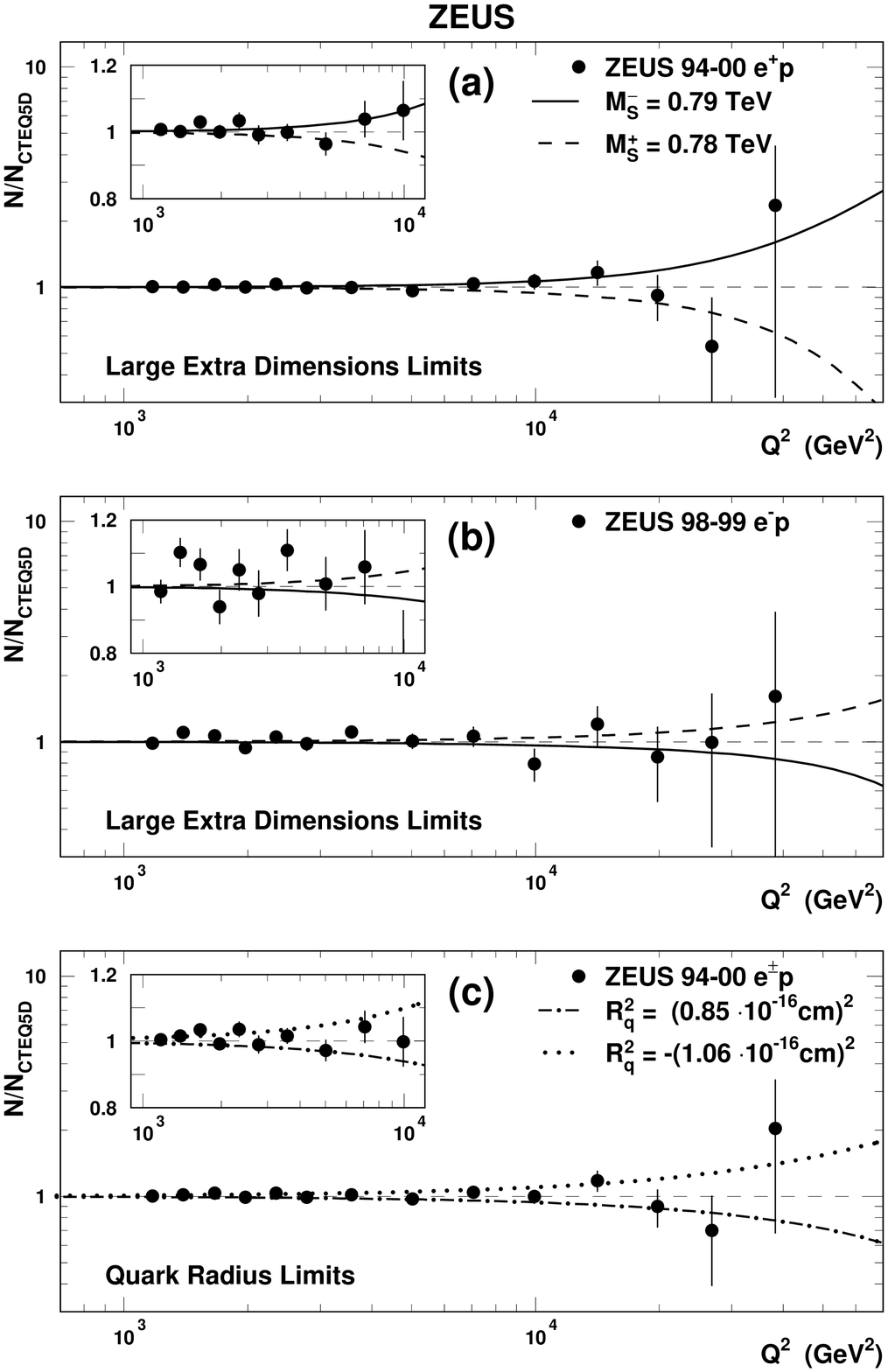,width=12cm}
\end{center}
\vspace{-0.7cm}
  \caption{
         ZEUS $e^+p$ data (a) and $e^-p$ data (b)
  compared with \CL{95} exclusion limits for 
  the effective Planck mass scale in models with large extra 
  dimensions, for positive ($M_{S}^{+}$) and negative  ($M_{S}^{-}$) 
  couplings. (c) Combined 1994-2000 data 
   compared with \CL{95} exclusion limits for 
  the effective mean-square radius of the electroweak charge of the quark.
  Results are normalized to the Standard Model expectations
  calculated using the CTEQ5D parton distributions.
  The insets show the comparison in the $Q^2 < 10^4 \gev^2$ region,
  with a linear ordinate scale.
           }
  \label{fig-edrq}
\end{figure}

%% file: DESY-03-218.bbl
\providecommand{\etal}{et al.\xspace}
\providecommand{\coll}{Coll.\xspace}
\catcode`\@=11
\def\@bibitem#1{%
\ifmc@bstsupport
  \mc@iftail{#1}%
    {;\newline\ignorespaces}%
    {\ifmc@first\else.\fi\orig@bibitem{#1}}
  \mc@firstfalse
\else
  \mc@iftail{#1}%
    {\ignorespaces}%
    {\orig@bibitem{#1}}%
\fi}%
\catcode`\@=12
\begin{mcbibliography}{10}

\bibitem{epj:c14:239}
ZEUS \coll, J.~Breitweg \etal,
\newblock Eur.\ Phys.\ J.{} {\bf C~14},~239~(2000)\relax
\relax
\bibitem{hab-91-01}
P.~Haberl, F.~Schrempp and H.~U.~Martyn,
\newblock {\em {\it Proc.\ Workshop on Physics at HERA}}, W.~Buchm\"uller and
  G.~Ingelman~(eds.), p.~1133.
\newblock Hamburg, Germany (1991)\relax
\relax
\bibitem{science:275:1759}
C.S.~Wood et al.,
\newblock Science{} {\bf 275},~1759~(1997)\relax
\relax
\bibitem{pr:d45:1602}
S.A.~Blundell, J.~Sapirstein and W.R.~Johnson,
\newblock Phys. Rev.{} {\bf D~45},~1602~(1992)\relax
\relax
\bibitem{prl:82:2484}
S.C.~Bennett and C.E.~Wieman,
\newblock Phys. Rev. Lett.{} {\bf 82},~2484~(1999)\relax
\relax
\bibitem{prl:85:1618}
A.~Derevianko,
\newblock Phys. Rev. Lett.{} {\bf 85},~1618~(2000)\relax
\relax
\bibitem{prl:86:3260}
M.G.~Kozlov, S.G.~Porsev, I.I.~Tupitsyn,
\newblock Phys. Rev. Lett.{} {\bf 86},~3260~(2001)\relax
\relax
\bibitem{pl:b191:442}
W.~Buchm\"uller, R.~R\"uckl and D.~Wyler,
\newblock Phys.\ Lett.{} {\bf B~191},~442~(1987).
\newblock Erratum in Phys.~Lett.~{\bf B~448}, 320 (1999)\relax
\relax
\bibitem{zfp:c46:679}
A.~Djouadi \etal,
\newblock Z.\ Phys.{} {\bf C~46},~679~(1990)\relax
\relax
\bibitem{pr:d68:052004}
ZEUS Coll., S.~Chekanov et al.,
\newblock Phys. Rev.{} {\bf D~68},~052004~(2003)\relax
\relax
\bibitem{zfp:c74:595}
J.~Kalinowski \etal,
\newblock Z.\ Phys.{} {\bf C~74},~595~(1997)\relax
\relax
\bibitem{pl:b429:263}
N.~Arkani-Hamed, S.~Dimopoulos and G.~Dvali,
\newblock Phys.~Lett.{} {\bf B~429},~263~(1998)\relax
\relax
\bibitem{pl:b436:257}
I.~Antoniadis et al.,
\newblock Phys. Lett.{} {\bf B 436},~257~(1998)\relax
\relax
\bibitem{pr:d59:086004}
N.~Arkani-Hamed, S.~Dimopoulos and G.~Dvali,
\newblock Phys.~Rev.{} {\bf D~59},~086004~(1999)\relax
\relax
\bibitem{np:b544:3}
G.F.~Giudice, R.~Rattazzi and J.D.~Wells,
\newblock Nucl. Phys.{} {\bf B 544},~3~(1999)\relax
\relax
\bibitem{pl:b460:383}
K.~Cheung,
\newblock Phys. Lett.{} {\bf B460},~383~(1999)\relax
\relax
\bibitem{pl:b479:358}
H1 \coll, C.~Adloff \etal,
\newblock Phys.\ Lett.{} {\bf B~479},~358~(2000)\relax
\relax
\bibitem{epj:c11:427}
ZEUS \coll, J.~Breitweg \etal,
\newblock Eur.\ Phys.\ J.{} {\bf C~11},~427~(1999)\relax
\relax
\bibitem{epj:c28:175}
ZEUS Coll., S.~Chekanov et al.,
\newblock Eur. Phys. J.{} {\bf C 28},~175~(2003)\relax
\relax
\bibitem{zeus:xsec}
ZEUS Coll., S.~Chekanov et al.,
\newblock {\em High-$Q^2$ neutral current cross sections in $e^+ p$ deep
  inelastic scattering at $\sqrt{s} = 318\gev$}.
\newblock Preprint DESY-03-214 (hep-ex/0401003), (2003). Submitted to Phys.
  Rev. D\relax
\relax
\bibitem{cpc:69:155}
A.~Kwiatkowski, H.~Spiesberger and H.-J.~M\"ohring,
\newblock Comp.\ Phys.\ Comm.{} {\bf 69},~155~(1992).
\newblock Also in {\it Proc.\ Workshop on Physics at HERA}, W.~Buchm\"uller and
  G.~Ingelman (eds.), p. 1294. Hamburg, Germany (1991)\relax
\relax
\bibitem{cpc:81:381}
K.~Charchu{\l}a, G.A.~Schuler and H.~Spiesberger,
\newblock Comp.\ Phys.\ Comm.{} {\bf 81},~381~(1994)\relax
\relax
\bibitem{spi:www:djangoh11}
H.~Spiesberger,
\newblock {\em {\sc heracles} and {\sc djangoh}: Event Generation for $ep$
  Interactions at {HERA} Including Radiative Processes}, 1998,
\newblock available on \texttt{http://www.desy.de/\til
  hspiesb/djangoh.html}\relax
\relax
\bibitem{cpc:71:15}
L.~L\"onnblad,
\newblock Comp.\ Phys.\ Comm.{} {\bf 71},~15~(1992)\relax
\relax
\bibitem{tech:cern-dd-ee-84-1}
R.~Brun et al.,
\newblock {\em {\sc geant3}},
\newblock Technical Report CERN-DD/EE/84-1, CERN, 1987\relax
\relax
\bibitem{epj:c12:375}
CTEQ \coll, H.L.~Lai \etal,
\newblock Eur.\ Phys.\ J.{} {\bf C~12},~375~(2000)\relax
\relax
\bibitem{pr:d55:1280}
H.L.~Lai \etal,
\newblock Phys.\ Rev.{} {\bf D~55},~1280~(1997)\relax
\relax
\bibitem{pr:d67:012007}
ZEUS \coll, S.~Chekanov \etal,
\newblock Phys.\ Rev.{} {\bf D~67},~012007~(2003)\relax
\relax
\bibitem{epdflib}
M. Botje,
\newblock {\em Fast access to parton densities, errors and correlations,
  EPDFLIB v. 2.0}.
\newblock NIKHEF-99-034\relax
\relax
\bibitem{epj:c14:285}
M.~Botje,
\newblock Eur.\ Phys.\ J.{} {\bf C~14},~285~(2000)\relax
\relax
\bibitem{pr:d66:010001}
Particle Data Group, K.~Hagiwara \etal,
\newblock Phys.\ Rev.{} {\bf D~66},~010001~(2002)\relax
\relax
\bibitem{epj:c21:443}
ZEUS \coll, S.~Chekanov \etal,
\newblock Eur.\ Phys.\ J.{} {\bf C~21},~443~(2001)\relax
\relax
\bibitem{pl:b539:197}
ZEUS \coll, S.~Chekanov \etal,
\newblock Phys.\ Lett.{} {\bf B~539},~197~(2002)\relax
\relax
\bibitem{pl:b568:35}
H1 Coll., C. Adloff et al.,
\newblock Phys. Lett.{} {\bf B568},~35~(2003)\relax
\relax
\bibitem{prl:79:2198}
CDF Coll., F.~Abe et al.,
\newblock Phys.\ Rev.\ Lett.{} {\bf 79},~2198~(1997)\relax
\relax
\bibitem{prl:82:4769}
\DO \coll, B.~Abbott \etal,
\newblock Phys.\ Rev.\ Lett.{} {\bf 82},~4769~(1999)\relax
\relax
\bibitem{epj:c12:183}
ALEPH Coll., R.~Barate et al.,
\newblock Eur.\ Phys.\ J.{} {\bf C~12},~183~(2000)\relax
\relax
\bibitem{epj:c11:383}
DELPHI Coll., P.~Abreu et al.,
\newblock Eur.\ Phys.\ J.{} {\bf C~11},~383~(1999)\relax
\relax
\bibitem{pl:b489:81}
L3 Coll., M.~Acciarri et al.,
\newblock Phys.\ Lett.{} {\bf B~489},~81~(2000)\relax
\relax
\bibitem{epj:c13:553}
OPAL Coll., G.~Abbiendi et al.,
\newblock Eur.\ Phys.\ J.{} {\bf C~13},~553~(2000)\relax
\relax
\bibitem{pl:b470:281}
L3 Coll., M.~Acciarri et al.,
\newblock Phys.\ Lett.{} {\bf B~470},~281~(1999)\relax
\relax
\bibitem{prl:86:1156}
\DO Coll., D.~Abbott et al.,
\newblock Phys.\ Rev.\ Lett.{} {\bf 86},~1156~(2001)\relax
\relax
\end{mcbibliography}
